\documentstyle[psfig]{mn}

\def\spose#1{\hbox to 0pt{#1\hss}}
\newcommand{\approxlt}{\mathrel{\spose{\lower 3pt\hbox{$\sim$}}
	\raise 2.0pt\hbox{$<$}}}
\newcommand{\approxgt}{\mathrel{\spose{\lower 3pt\hbox{$\sim$}}
	\raise 2.0pt\hbox{$>$}}}

\newcommand {\ergcms}        {{\rm erg\,cm^{-2}\,s^{-1}}}
\newcommand {\hmpc}          {\,h^{-1}\,{\rm Mpc}}
\newcommand {\rc}            {r_{\rm c}}

\newcommand {\nobs}          {N_{\rm obs}}
\newcommand {\npois}         {N_{\rm P}}
\newcommand {\nexp}          {N_{\rm exp}}

\newcommand{\ROSAT}{{\sl ROSAT}}

\begin{document}

\title[Clustering of X--ray selected AGN]
{Clustering of X--ray selected Active Galactic Nuclei}

\author[F.J. Carrera, X. Barcons, A.C. Fabian et al.]
	{ F.J. Carrera$^{1,2}$, X. Barcons$^{1,3}$, A.C. Fabian$^{3}$,
	  G. Hasinger$^{4}$, K.O. Mason$^{2}$, \and
	  R.G. McMahon$^{3}$, J.P.D. Mittaz$^{2}$ and  M.J. Page$^{2}$
	\\
	$^1$ Instituto de F\'{\i}sica de Cantabria (Consejo Superior
	de Investigaciones Cient\'{\i}ficas - Universidad de Cantabria),\\
        {\,\,\,\,}Avenida de los Castros, 39005 Santander, Spain\\
	$^2$ Mullard Space Science Laboratory--University College London,
	Holmbury St. Mary, Dorking, Surrey RH5 6NT\\
	$^3$ Institute of Astronomy, Madingley Road, Cambridge CB3 0HA\\
	$^4$ Astrophysikalisches Institut Potsdam, An der Sternwarte 16,
        Potsdam, Germany\\
	}

\maketitle

\begin{abstract}      

A total of 235 Active Galactic Nuclei (AGN) from two different soft
X-ray surveys (the \ROSAT\ Deep Survey --DRS-- and the \ROSAT\
International X-ray Optical Survey --RIXOS--) with redshifts between 0
and 3.5 are used to study the clustering of X-ray selected AGN and its
evolution.  A 2~sigma significant detection of clustering of such
objects is found on scales $<40-80\hmpc$ in the RIXOS sample, while no
clustering is detected on any scales in the DRS sample.
Assuming a single power law model for the spatial correlation function
(SCF), quantitative limits on the AGN clustering have been obtained: a
comoving correlation length $1.5\approxlt r_0 \approxlt 3.3\hmpc$ is
implied for comoving evolution, while $1.9\approxlt r_0 \approxlt 4.8$
for stable clustering and $2.2\approxlt r_0 \approxlt 5.5$ for linear
evolution.  These values are consistent with the
correlation lengths and evolutions obtained for galaxy samples, but 
imply smaller amplitude or faster evolution than recent UV and optically
selected AGN samples.  We also constrain the ratio of bias
parameters between X-ray selected AGN and $IRAS$ galaxies to be
$\approxlt 1.7$ on scales $\approxlt 10\hmpc$, a significantly smaller
value than is inferred from local large-scale dynamical studies.
 
\end{abstract}


\section{INTRODUCTION}

Since the launch of \ROSAT\ in 1991, a large number of surveys of soft
X--ray selected sources have been undertaken using the Position
Sensitive Proportional Counter (PSPC), with different sky coverages,
depths and identification completenesses.  They range from the \ROSAT\
All Sky Survey (RASS, Voges 1992) that covers the whole sky to a fairly
bright flux limit, passing through the relatively large solid angle RIXOS (20
$\deg^2$ Mason et al., in preparation), to the deeper \ROSAT\ Deep Survey
(henceforth DRS, Shanks et
al. 1991), Cambridge Cambridge \ROSAT\ Serendipitous Survey (Boyle et
al. 1995) or the UK Medium Survey (Carballo et al. 1995). The deepest
and narrowest surveys, pushing to the limit the capabilities of the
PSPC are the UK Deep Survey (Branduardi--Raymont et al. 1994, McHardy
et al. 1998) and the Deep Survey in the Lockman Hole (Hasinger et
al. 1993, Hasinger et al. 1998, Schmidt et al. 1998).

Together, these surveys have given considerable insight into the
evolution of the AGN X--ray luminosity function (Boyle et al. 1994,
Page et al.  1996, Jones et al. 1997), their X--ray spectra (Bade et
al. 1995, Carballo et al. 1995, Almaini et al. 1996, Romero--Colmenero
et al. 1996,  Ciliegi et al. 1997, Mittaz et al. 1998) and their
relation to Narrow Emission Line Galaxies (NELGs, Boyle et al. 1995,
Griffiths et al. 1996, Page et al. 1997).

However, so far the only direct study of the clustering properties of
X--ray selected AGN is that by Boyle \& Mo (1993, henceforth BM).
They studied the local ($z<0.2$) AGN in the Extended Einstein Medium
Sensitivity Survey (EMSS, Stocke et al. 1991) and found no signal
above 1 sigma, consistent with the clustering properties of the UV and
optically selected AGN.

In this work we present for the first time a direct study of the
clustering of soft X--ray selected AGN from \ROSAT, using RIXOS and
the \ROSAT\ Deep Survey (positions and redshifts for the AGN
identified in the five DRS fields have been kindly provided by
O. Almaini, prior to publication).

The samples are introduced in Section \ref{data}. We discuss the
evidence for clustering in Section \ref{clust}, using different model
independent methods. Quantitative measurements on the correlation
strength and evolution are then obtained in Section
\ref{calscf}. These results are presented in Section \ref{limits}, and
discussed in Section \ref{discuss}. We summarize our conclusions in Section
\ref{summa}.

We have used $H_0=100\,h$ km s$^{-1}$ Mpc$^{-1}$ and $q_0=0.5$ (unless
otherwise stated) throughout this paper.

\section{THE DATA}
\label{data}

We have used two large complete soft X--ray AGN samples: DRS (Boyle et
al. 1994, Shanks et al., in preparation), and RIXOS (Page et al. 1996,
Mason et al., in preparation).  

\begin{figure}
\vbox to 0cm{\vfil}
{\psfig{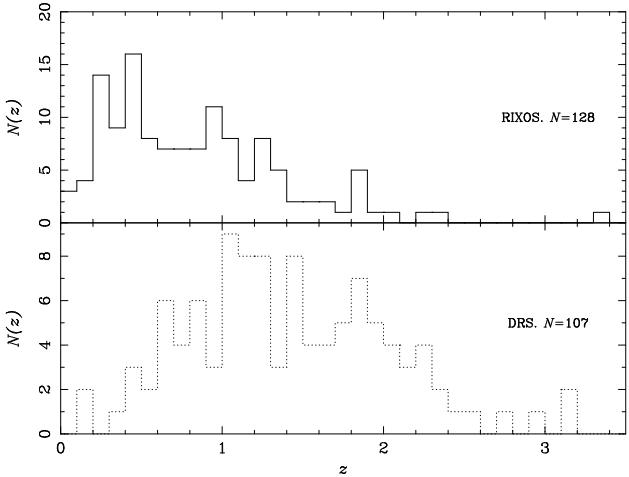}}
\caption{Histogram of the distribution of the redshifts of the AGN in RIXOS
and DRS (see text).}
\label{Fig1}
\end{figure}

The DRS sample is the deepest, being a `pencil beam' style survey in a few
chosen directions in the sky, while RIXOS is wider (a collection of
many shallower pencil beams), being a compromise
between depth and surveyed solid angle. Both surveys together cover a wide
range of redshifts (see Fig. \ref{Fig1}) and allow an investigation of
the clustering properties of X--ray AGN.

\begin{description}

\item {\bf DRS}: Five deep PSPC exposures have been searched for
sources down to the sensitivity limit of that instrument over the
inner field of view (off--axis angle $<$18~arcmin). Due to the
increasing width of the Point Spread Function (PSF) with off--axis
radius and to vignetting, the flux limit of this survey is a function
of off--axis radius and different for each field. This has been taken
into account in our study, as well as the fact that the source
identification completeness also varies with flux (see Section
\ref{pairs}).  A total of 107 X--ray AGN have been detected in these
fields, with redshifts between 0 and 3.5.

\item {\bf RIXOS}: This is a wider solid angle shallower survey than the DRS,
in which the central 17~arcmin radius solid angle of 80 medium--deep PSPC
pointings with $\mid b \mid > 20^\circ$ and exposure times longer than
8000 s have been source searched.  About 90 per cent of the sources in
a total of 65 fields have been identified down to a uniform limit of
$3\times 10^{-14}\ergcms$ (0.5--2~keV), significantly higher than the
sensitivity limit of all fields (Mason et al., in preparation). We
have further selected those fields with an exposure time longer than
10000 s, resulting in 43 fields (see Table \ref{Table0}) and 128 AGN
with redshifts between 0 and 3.5.

\end{description}

\begin{table}
\centering
%  \vbox to 0cm{\vfil} 
% \begin{minipage}{22cm} 
\caption{RIXOS fields used. The column labelled `RIXOS' gives the RIXOS Field
identification number and the column labelled `\#' is the \ROSAT\ sequence
identification number (see Section \protect\ref{data}).}
\label{Table0} 
\begin{tabular}{ c c c c}
RIXOS & \# & RIXOS & \# \\
\hline
 110 & 200329rp &  234 & 700112rp \\
 115 & 000049rp &  240 & 700055rp \\
 116 & 000054rp &  245 & 700099rp \\
 122 & 170174rp &  248 & 700329rp \\
 123 & 700228rp &  252 & 700319rp \\
 125 & 200322rp &  253 & 700387rp \\
 126 & 700223rp &  254 & 700391rp \\
 205 & 100578rp &  255 & 700315rp \\
 206 & 200453rp &  257 & 700326rp \\
 211 & 700210rp &  258 & 700358rp \\
 215 & 150046rp &  259 & 700010rp \\
 216 & 700211rp &  260 & 300158rp \\
 217 & 700248rp &  261 & 201103rp \\
 219 & 700208rp &  262 & 701048rp \\
 220 & 701200rp &  265 & 700216rp \\
 221 & 700546rp &  268 & 700392rp \\
 223 & 200721rp &  271 & 700510rp \\
 224 & 100308rp &  272 & 700489rp \\
 225 & 200076rp &  273 & 700384rp \\
 226 & 700073rp &  274 & 700227rp \\
 227 & 200091rp &  302 & 700540rp \\
 228 & 400020rp \\
\hline
\end{tabular}

% \end{minipage}
\end{table}

\begin{table}
\centering
%  \vbox to 0cm{\vfil} 
% \begin{minipage}{22cm} 
\caption{The different samples. `$N$' are the number of AGN in each sample.
The column labelled `\%' are the percentages of Poisson simulations with
likelihood values higher than those of each sample (see Section
\protect\ref{MLcorrel}). The column labelled `$D_c$' gives an estimate
of the mean comoving distance between the objects in each sample
(see Section \protect\ref{data}). The samples
labelled with an asterisk are the
ones used in Sections \protect\ref{calscf} and \protect\ref{limits}.}
\label{Table1} 
\begin{tabular}{ l c c r r c }
Survey & $z$ interval & $\langle z \rangle$ & $N$ & \%  & $D_c$ ($\hmpc$)\\
\hline
RIXOS*        & 0.0-3.5 & 0.838 & 128 & 61 & 362 \\
RIXOS         & 0.0-0.5 & 0.321 &  46 & 42 & 513 \\
RIXOS         & 0.5-1.0 & 0.768 &  40 &  4 & 235 \\
RIXOS         & 1.0-2.0 & 1.362 &  38 & 51 &  -  \\
RIXOS         & 0.0-1.0 & 0.529 &  86 & 17 & 288 \\
              &         &       &     &    &     \\
DRS*          & 0.0-3.5 & 1.425 & 107 & -  &\ 49 \\
DRS           & 0.0-1.0 & 0.668 &  27 & -  & 161 \\
DRS           & 1.0-3.5 & 1.680 &  80 & -  &\ 38 \\
\hline
\end{tabular}

% \end{minipage}
\end{table}

The median luminosities of the RIXOS and DRS samples are similar
$L_{\rm 0.5-2\,keV}\sim 0.2\times10^{44}\,{\rm erg\,s^{-1}}$ (assuming
a power law spectrum with an $\alpha=1$ energy index). The average
values are however different $\langle L_{44}\rangle\sim0.46$ for RIXOS
while $\langle L_{44}\rangle\sim0.23$ for DRS. This indicates that,
while the sources in both samples have similar overall luminosities,
the luminous AGN in RIXOS are brighter than those in the DRS (see Figs. 2 of
Boyle et al. 1994 and Page et al. 1996). We will see that this does
not have any effect on our results in Section \ref{pairs}.

The number of objects in each sample and the average redshifts for
different redshift ranges are shown in Table \ref{Table1}. Also given
are $D_c$ the comoving distances below which the total number of observed pairs
(see Section \ref{pairs}) equals half the number of objects in the
sample. This is an estimate of the mean distance between the objects
in each sample, given the `pencil beam'  geometry of
each of the \ROSAT\ fields that make up our two surveys. $D_c$ is
missing for the RIXOS $1<z<2$ sample because there are only 16 pairs in
total in that sample. Our default
samples are those marked with an asterisk in Table \ref{Table1}: the
total RIXOS and DRS samples, a total of 235 AGN. These two samples
have complementary redshift distributions: most RIXOS AGN are
at $z<1$ ($\langle z\rangle=0.84$), while most DRS AGN
are at $z>1$  ($\langle z\rangle=1.43$) (see Fig. \ref{Fig1} and Table
\ref{Table1}). In this sense, RIXOS reflects the `local' behaviour of
AGN clustering, while DRS can constrain its `medium--high redshift' evolution.

\section{Is there any evidence for clustering in these samples?}
\label{clust}

We have investigated the presence of clustering in the RIXOS sample in
two different model--independent ways: by comparing the distribution
of the number of sources per field in each sample with that expected
from a purely Poisson distribution (a variant of the counts--in--cells
method, but in angular rather than spatial cells) and by comparing the
total number of pairs of sources separated by a comoving distance
$\rc$ with that expected from a uniform spatial distribution of
sources. This second method has also been used for the DRS sample, the
first not being adequate due to the small number of fields (5) of that
sample.

\subsection{Counts--in--cells: checking the uniformity of our RIXOS sample}
\label{MLcorrel}

If $N$ is the total number of AGN in a sample, $n$ is the total number
of fields ($n=43$ for RIXOS), $\mu\equiv N/n$ is the observed average number
of sources per field, and $N_i$ is the number of sources in 
field $i$, we define a likelihood

\begin{equation}
{\rm L}=-\sum_i \log{P_\mu (N_i)}
\end{equation}

\noindent where $P_\mu (N_i)$ is the Poisson probability of finding $N_i$ from
a Poisson process of average $\mu$

\begin{equation}
P_\mu (N)={ \mu^N e^{-\mu} \over N! }
\end{equation}

For each sample, 1000 Poisson simulations with the same $\mu$, $N$ and
$n$ as the real sample are performed, and the likelihood L is
calculated for each one of them.  The percentage of Poisson
simulations with likelihood values larger than those of the
corresponding observed samples are given in the last column of Table
\ref{Table1}.

We can see that 96 per cent of the Poisson simulations have `better'
likelihood than the RIXOS $0.5<z<1$ sample or, in other words, there
is evidence for clustering at the 2 sigma level in RIXOS in the 0.5-1
redshift interval. This is independent of the nature of the clustering.

The significance becomes smaller if we consider together all AGN with
$0<z<1$, probably because the volume sampled at low redshift is much
smaller. The lack of clustering signal in the higher redshift bin
($1<z<2$) is probably due to a combination of the falling sensitivity
(typical in a flux--limited survey) and the lower clustering amplitude
at higher redshift (`positive evolution' see Section \ref{limits}).
This method does not detect any significant clustering in the whole
RIXOS sample ($0<z<3.5$).

This test has also been used to check for possible effects of the
galactic absorption and/or exposure times on the mean density of
sources in different fields. We have found that there is not any
significant difference between the low and high column density fields,
nor between the shorter and longer exposure time fields, in terms of
the surface density of RIXOS sources. This test was also repeated for
the faint and bright sources separately, finding the same negative
result.  We can thus be confident in the uniformity of our source
sampling with respect to `instrumental' selection effects.

\subsection{Pairs of sources}
\label{pairs}

The previous method discards all the information on the spatial separation of
the sources. It is obvious that two sources in the same field, but at
the opposite ends of the redshift interval, are physically unrelated.
The counts--in--cells method used above does not have a way of
discriminating against such cases, unless the redshift intervals are
made smaller in which case the quality of the statistics  worsens.

We have performed new simulations in which as many sources as in the
real samples are redistributed at random among the different fields
(but not between the two different samples), keeping their redshifts
and fluxes, but randomizing their off--axis angles and `azimuths'
(angle between the line joining the field centre to the source and the
meridian through the field centre).  This has been done in a different
way for the two samples:

\begin{description}

\item {\bf DRS}: In this sample, the survey effective solid angle was
different for every field and a function of off--axis angle within
every field (see Table 2 of Boyle et al. 1994). A total of 25
different regions were defined, 5 for every field, corresponding to
the off--axis ranges 0--10~arcmin, 10--12~arcmin, 12--14~arcmin,
14--16~arcmin and 16--18~arcmin. The flux limit $S_{\rm lim}$ for
source detection was different in each one of these 25 regions, as was
the overall identification completeness for sources with $S>S_{\rm
lim}$. The `effective' solid angle $\Omega_{\rm eff}(S_{\rm lim})$ of
the survey for $S>S_{\rm lim}$ is defined so that $\Omega_{\rm
eff}(S_{\rm lim})/\Omega_{\rm geom}(S_{\rm lim})$ is the fraction of
identified sources down to that flux limit, where $\Omega_{\rm
geom}(S_{\rm lim})$ is the total solid angle surveyed, also down to
the same flux limit. When a simulated source with flux $S'$ was
extracted, we found the highest $S'_{\rm lim}$ to that $S'>S'_{\rm
lim}$. The source could be detected over the total surveyed area with
$S_{\rm lim}\leq S'_{\rm lim}$. A particular region within that area
(and hence a field) was assigned at random to the source,
proportionally to $\Omega_{\rm eff}(S_{\rm lim})/\Omega_{\rm
eff}(S'_{\rm lim})$. Finally, the off--axis angle and azimuth were
obtained assuming an uniform source distribution within the
corresponding off--axis range.

\item {\bf RIXOS}: The source density is uniform within every field
and field to field in this sample. Therefore, we assigned at random a
RIXOS field to every source, and an off--axis angle (within 17~arcmin)
and azimuth within that field.

\end{description}

The number $\nobs$ of pairs of sources in the two (separate) real samples with
comoving distance $\leq \rc$ is then obtained, as well as the mean
number (among 100000 simulations) of random pairs $\npois$ up to the
same separation.  The distances $\rc$ are calculated in comoving
coordinates according to the expressions given by Osmer (1981), for
$q_0=0$ and 0.5. Only pairs of sources in the same (real or simulated)
field are used, to facilitate the calculation of the volume integrals
introduced in Section \ref{calscf}.

We have used the integrated Poisson probability of obtaining
$<\nobs$ pairs for a distribution of mean $\mu$($=\npois$):

\begin{equation}\label{pnltn}
P_\mu (<\nobs)\equiv \sum_{N=0}^{\nobs-1} P_\mu(N)
\end{equation}

\begin{figure}
 \vbox to 0cm{\vfil}
{\psfig{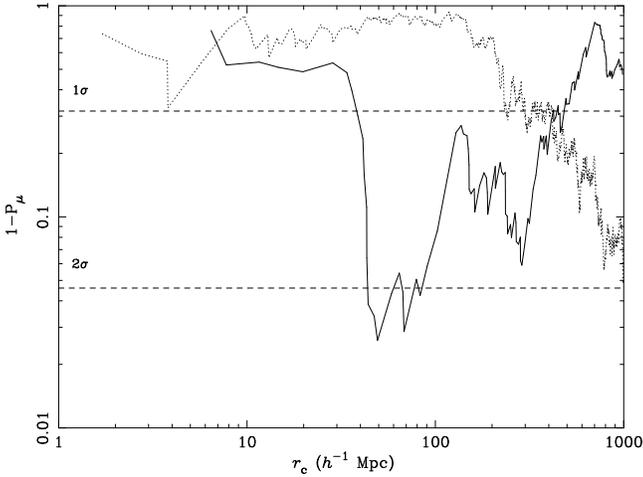}}
\caption{Complement to one of the significance of clustering as a
function of the comoving distance $\rc$, for the RIXOS (solid line)
and the DRS AGN (dotted line). 1 and 2~sigma significance levels
are shown with dashed lines}
\label{Fig2}
\end{figure}

The complements to one of these probabilities (i.e. $1-P_\mu$) are
plotted in Fig. \ref{Fig2}, for the $\rc$ values corresponding to each
real pair in the RIXOS sample and the DRS sample. It is
clear from this plot that there is a detection of clustering (at the
$\sim$95 per cent or 2~sigma level) in the RIXOS sample, at
comoving distances $\rc\approxlt 40-80\hmpc$ (e.g., $\nobs=22$ pairs
observed for $\npois=14.6$ pairs expected for $\rc\leq
83.03\hmpc$). We have checked that the sources contributing to this
signal are unrelated to the targets of the corresponding \ROSAT\
pointings. As a general comment, the targets are mostly at $z<0.2$
while the sources contributing to the signal have $0.2<z<1.4$.

It is also clear that there is not a significant detection of
clustering in the DRS sample at any separation (e.g., for $\rc\leq
5.68\hmpc$, $\nobs=5$ and $\npois=4.85$). None of these results
changes significantly if $q_0=0$ is used instead of the default
$q_0=0.5$.

We also note the lack of significant clustering at the smallest
separation of any RIXOS pairs at ($\rc\leq 6.44\hmpc$, $\nobs=1$ and
$\npois=1.44$). This fact, along with the non--detection in the DRS sample
at similar scales and the detection of clustering at $\rc\leq 80\hmpc$
will be used in the next section to constrain the clustering amplitude
for different clustering evolution models.

The lack of very close pairs is not due to the limited angular
resolution of \ROSAT. We have repeated the calculation of both the
real and simulated pairs excluding those pairs with an angular
distance smaller than 1~arcmin (larger than the minimum distance at
which the \ROSAT-PSPC could resolve two separate sources at the flux
levels relevant here), finding very similar results.

We have investigated if the detection of clustering in RIXOS and not
in the DRS could be due to the higher luminosity of the RIXOS sources.
The average luminosity of the RIXOS sources contributing to the
$\rc<100\hmpc$ pairs is $\langle L_{44}\rangle=0.25$, while that
of the DRS sources at the same separations is $\langle L_{44}\rangle
=0.24$, the median luminosities being $L_{44}=0.17$ and
$L_{44}=0.19$, respectively. Clearly, there is not any significant
difference in luminosities between the two samples to which the
observed difference in clustering strength could be
attributed. However, there is a difference in their redshift
distributions: all of the RIXOS sources at those separations are at
$z<1.4$, while most of the DRS sources have $z>1$.  This is not very
surprising, considering that most models of structure formation
predict the growth of inhomogeneities with cosmic time, or less
clustering at higher redshift. 

Finally, we would like to emphasize
that this test is sensitive to all pairs with comoving separations
smaller or equal to the value of $\rc$ shown. It is also important to
stress the independence of this test on any particular clustering
model.

\section{The integrated spatial correlation function}
\label{calscf}

Since there is an excess of AGN pairs up to a certain comoving
separation, with respect to the expectation from a uniform
distribution, we include now the possible spatial clustering of the
sources.  Indeed, the expected number of pairs up to a certain
comoving separation will depend on the source correlation function
integrated up to the corresponding comoving separation and taking into
account the detailed geometry of the volumes sampled by the RIXOS and
DRS observations.

\begin{figure}
 \vbox to 0cm{\vfil}
{\psfig{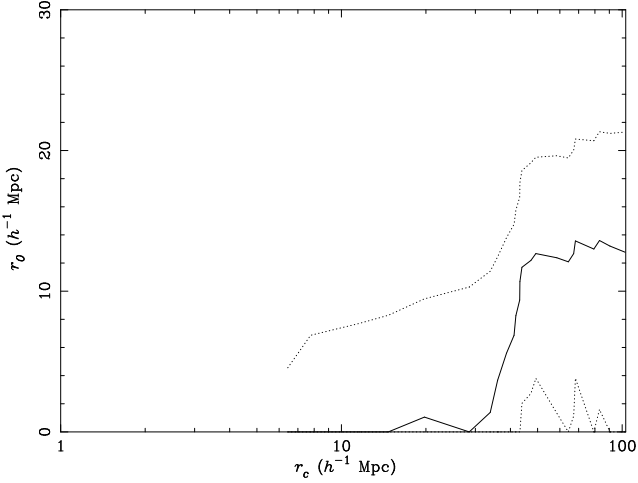}}
\caption{The `best fit' $r_0$ (solid line) as a function of the
comoving separation $\rc$ for stable clustering for the RIXOS
AGN, as well as the 2~sigma limits (dotted lines).}
\label{Fig3}
\end{figure}

The spatial (3-dimensional) correlation function $\xi(r)$ is defined
as an excess probability (Peebles 1980). If $n$ is the spatial density
of the sources under study, the probability of finding a source in the
volume $dV_1$ and another one in $dV_2$, separated by $r_{12}$, is:

\begin{equation}
\delta P = n^2 (1+\xi(r_{12}))dV_1 dV_2
\end{equation}
As mentioned above, the relevant quantity is the integrated spatial
correlation function (SCF):

\begin{equation}
\bar \xi (\rc)= {1\over V} \int_{V} dV \xi(r)
\end{equation}
where $V$ is the (comoving) volume over which the pairs are counted. In the
present paper, that volume is the intersection of a sphere of radius
$\rc$ with the cone defined by the maximum off--axis angles
of each one of our \ROSAT\ fields and the redshift limits of our samples.
We call $V_i$ to such volume around each of our sources with redshift
$z_i$.

For the DRS sample, this is further refined by slicing this sphere
along the conical shells defined for the different effective solid
angles and flux limits for every field (see Section \ref{pairs}), and
summing up the integrals over those shells weighting them by

\begin{equation}
f= { N(>S_{\rm lim})\over N(>S_{\rm min})} 
   \times
   { \Omega_{\rm eff}(S_{\rm lim}) \over \Omega_{\rm geom}(S_{\rm lim}) }
\end{equation}

\noindent where $N(>S)$ is the surface density of sources with flux
greater or equal to $S$, $S_{\rm lim}$ is the flux limit of the
corresponding off--axis shell, $S_{\rm min}$ is the lowest flux limit
of the DRS ($S_{\rm 0.5-2\,keV}=0.32\times 10^{-14}\ergcms$),
$\Omega_{\rm eff}(S_{\rm lim})$ is the effective solid angle at
$S_{\rm lim}$, and $\Omega_{\rm geom}(S_{\rm lim})$ is the total
surveyed solid angle at $S_{\rm lim}$. This takes into account the
fraction of AGN missed by the DRS due to the different flux limits at
different off--axis radii, and the fraction missed due to the
flux--dependent identification incompleteness of that survey.

The expected mean number of pairs in a volume $V$ is given by
(Peebles 1980):

\begin{equation}
\nexp = \npois (1+\bar\xi(\rc))
\end{equation}

Given a model for the SCF, the above integrals can be performed,
and the parameters of the model constrained using $P_{N{\rm exp}} (<\nobs)$
(see equation \ref{pnltn}). For example, a model with an amplitude too
large would produce a value of $\nexp$ too high compared to $\nobs$
that would then be highly unlikely.

We have assumed the usual power--law shape for the SCF

\begin{equation}
\xi(\rc,z)=(1+z)^{-p}(\rc/r_0)^{-\gamma}
\end{equation}

\noindent where $r_0$ is the correlation length, $\gamma=1.8$ (the
results presented below do not change significantly if $\gamma=1.6$ is
used instead) and $p$ is an evolutionary parameter (BM). If $p=0$ the
clustering is constant in comoving coordinates (`comoving
evolution'). If the clustering is constant in physical coordinates
(`stable clustering') then $p=1.2$. Linear growth of structures
corresponds to $p=2$ in an $\Omega=1$ universe, higher values of $p$
representing even faster non--linear growth.

The evolution of the SCF is often parameterized, in physical coordinates, as

\begin{equation}
\xi(r,z)=(1+z)^{-(3+\epsilon)}(r/r_0)^{-\gamma}
\end{equation}

\noindent where $r$ is the physical distance and $\epsilon\equiv \gamma+p-3$
(Groth \& Peebles 1977).

Hence:

\begin{eqnarray}\label{nexppr0}
\lefteqn{ \nexp(p,r_0)=\sum_{\Delta z}\npois(\Delta z) \times } \nonumber\\
                  & & \times \left( 1 + {r_0^\gamma\over N_{z_i \in \Delta z}}
                  \sum_{z_i\in \Delta z}{(1+z_i)^{(-p)}\over V_i}
                  \int_{V_i} dV \rc^{-\gamma}\right)
\end{eqnarray}

\noindent where $\npois(\Delta z)$ is the number of poisson pairs
within a redshift interval $\Delta z$ ($\npois=\sum_{\Delta
z}\npois(\Delta z)$) from the simulations. The sum $\sum_{z_i}$ is
performed over all the sources with redshifts in $\Delta z$, and it is
essentially an average within $\Delta z$ of $\bar \xi(\rc)$ taking
into account the clustering evolution model assumed in each case. The
sum over $\Delta z$ is to deal with the changing spatial density of
our sources as a function of redshift, plus the geometry of our
surveys and their flux limits (see Section \ref{pairs}).

We show in Fig. \ref{Fig3} the `best fit' $r_0$ (the value of $r_0$
that makes $\nexp=\nobs$) and the 2~sigma limits on $r_0$ for stable
clustering ($p=1.2$), as a function of $\rc$ for the distances between
each real pair of RIXOS AGN (each new point in the lines uses the
cumulative distribution of pairs up to distance $\rc$). We can see
that there are only upper limits to the value of $r_0$ at small
separations, but the required value of $r_0$ becomes
different from zero at separations $\rc\approxlt
40-80\hmpc$, in agreement with the findings of Section \ref{pairs}.

Similarly, each pair $(p,r_0)$ can be assigned a probability for a
fixed value of $\rc$ using equations \ref{pnltn} and \ref{nexppr0}. 
This will be used in the next section to constrain the clustering
of X--ray selected AGN and/or its evolution.

In Figs.~2 and 3 we see that we have to go up to comoving separations
of $\sim 40\hmpc$ to find a significant excess of pairs with respect
to an unclustered population of sources.  Fig.~3 also shows that the
required comoving correlation length is much smaller, so the detection
of the clustering signal occurs mostly at the tails of the correlation
function where clustering is weak. We do not detect an excess of pairs
at comoving separations $r_c<10\hmpc$ probably because of the
relatively low density of objects in our sample, in which case the
number of expected pairs is always small at small separations.
Although all this might seem puzzling, Fig.~3 shows that a comoving
correlation length $r_0\approx 5\hmpc$ is consistent at all
separations. 

\section{LIMITS ON THE CLUSTERING STRENGTH FOR DIFFERENT EVOLUTION MODELS}
\label{limits}

\begin{figure}
 \vbox to 0cm{\vfil}
{\psfig{figure=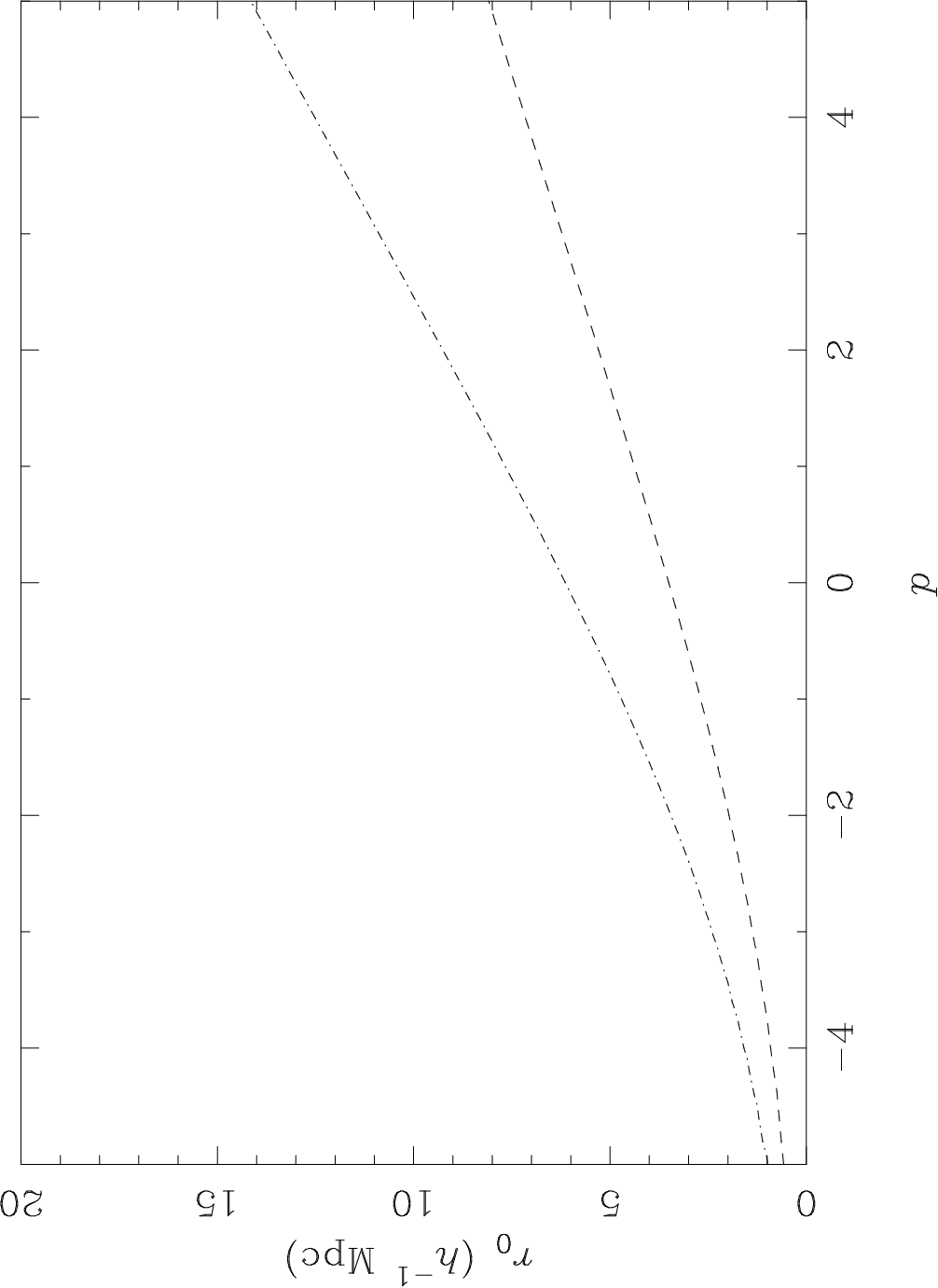,height=7truecm,bbllx=25pt,bblly=50pt,bburx=587pt,bbury=774pt,angle=270.0}}
\caption{Probability contour levels corresponding to 1~sigma (solid line), 
2~sigma (dashed line) and 3~sigma (dot--dashed line) in the $(p,r_0)$
space for the RIXOS AGN at distances $\rc\leq6.44\hmpc$.}
\label{Fig4}
\end{figure}

We show in Fig. \ref{Fig4} the 1, 2 and 3 sigma contours in the
$(p,r_0)$ space for the distance of the closest observed pair in the
RIXOS sample ($\rc\leq 6.44\hmpc$). Since no clustering is
detected at those separations, only upper limits to the value of $r_0$
for each $p$ can be obtained, ranging from $r_0\approxlt 3.8\hmpc$ for
comoving clustering to $r_0\approxlt 5.5\hmpc$ for linear growth (2 sigma).

\begin{figure}
 \vbox to 0cm{\vfil}
{\psfig{figure=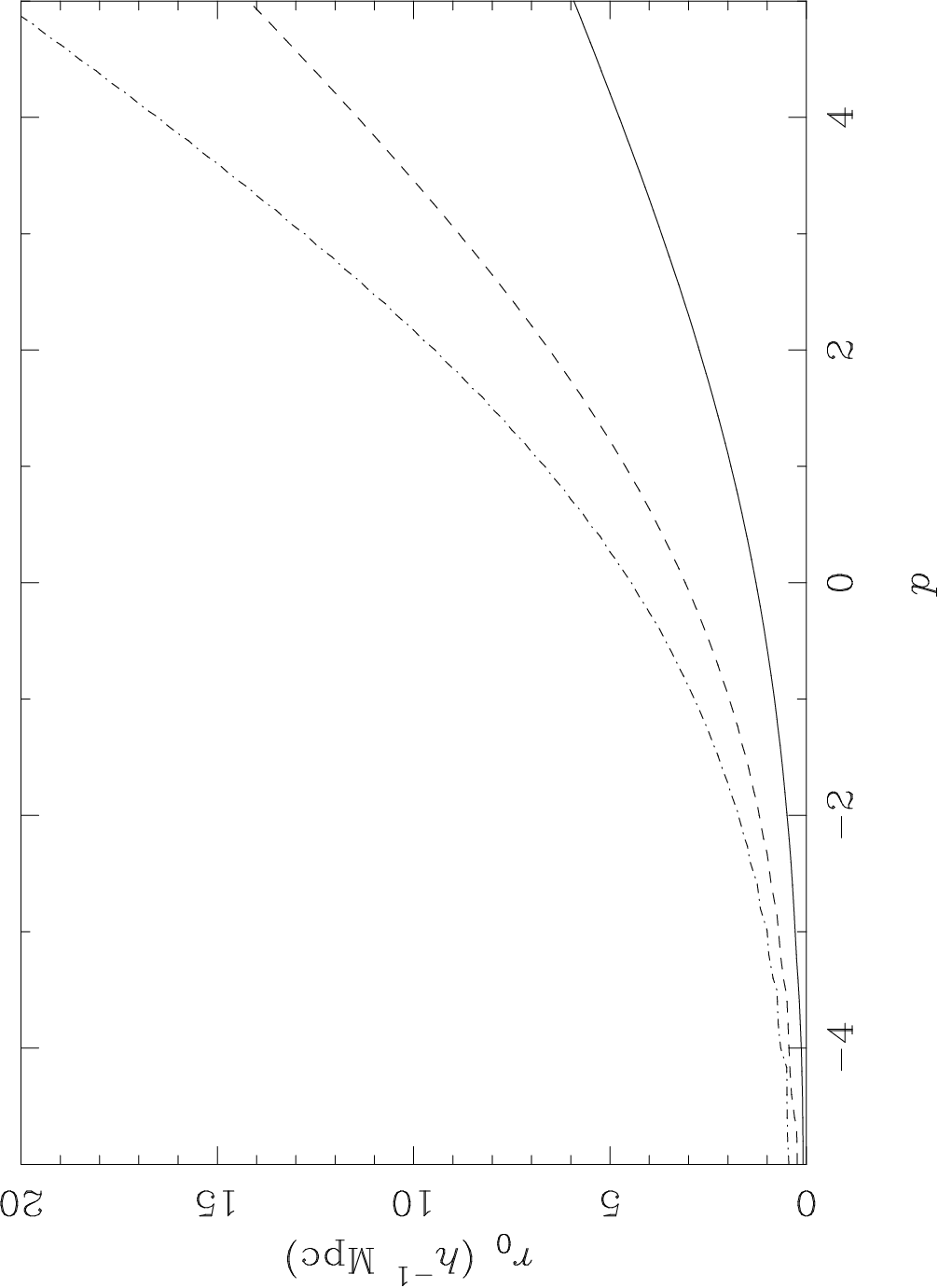,height=7truecm,bbllx=25pt,bblly=50pt,bburx=587pt,bbury=774pt,angle=270.0}}
\caption{Probability contour levels corresponding to 1~sigma (solid line),
2~sigma (dashed line) and 3~sigma (dashed--dotted line) in the $(p,r_0)$ space
for the DRS AGN at distances $\rc\leq 5.68\hmpc$.}
\label{Fig5}
\end{figure}

The 1, 2 and 3 sigma contours are plotted in Fig. \ref{Fig5} for the
DRS AGN at $\rc\leq 5.68\hmpc$. This distance has been chosen to
represent similar scales to those in the first RIXOS pair, but, due to
the absence of any signal of clustering in the DRS sample, it is
representative of the whole set of distances. Only upper limits are
found. They are more restrictive than those from RIXOS for $p\leq0$,
because DRS is deeper than RIXOS. Hence, it can constraint better the
high redshift behaviour of clustering ($p<0$ implies stronger
clustering at higher redshift).

\begin{figure}
 \vbox to 0cm{\vfil}
{\psfig{figure=fig6.ps,height=7truecm,bbllx=25pt,bblly=50pt,bburx=587pt,bbury=774pt,angle=270.0}}
\caption{Probability contour levels corresponding to 1~sigma (solid line),
2~sigma (dashed line) and 3~sigma (dashed--dotted line) in the $(p,r_0)$ space
for the RIXOS AGN at distances $\rc\leq 83.03\hmpc$.}
\label{Fig6}
\end{figure}

\begin{figure}
 \vbox to 0cm{\vfil}
{\psfig{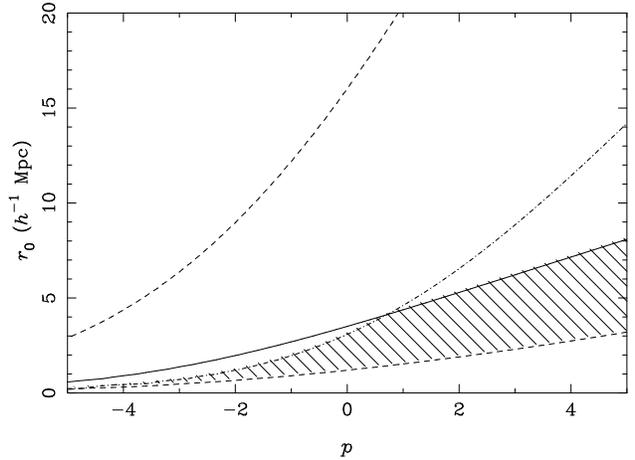}}
\caption{Probability contour levels corresponding to 2~sigma
 from RIXOS $\rc<6.44\hmpc$ (solid line) and
$\rc<83.03\hmpc$ (dashed lines), and DRS $\rc<5.68\hmpc$ (dashed--dotted
line).  The allowed region is that below the solid and dashed--dotted lines
and between the two dashed lines (shown shaded).}
\label{Fig7}
\end{figure}

The allowed region of the $(p,r_0)$ space from the clustering signal
in RIXOS is shown in Fig. \ref{Fig6}. The comoving distance at which this is
calculated ($\rc \leq 83.03\hmpc$) is the largest one below which there
is any clustering signal (at 2~sigma) in RIXOS, and is taken as a
representative value.  $r_0=0$ (no clustering) is excluded at
$>$2~sigma level for any value of $p$, in accordance with our
detection of clustering.

Figs. \ref{Fig4}, \ref{Fig5} and \ref{Fig6} can be combined to produce
a joint set of limits on the clustering strength and evolution, using
the RIXOS small and large scale `low redshift' results and the DRS small
scale `high redshift' results.  The 2~sigma contours from those three
plots are shown in Fig. \ref{Fig7} together. Note that a region in the
$(p,r_0)$ parameter space `outside' any of the 2~sigma contours is
actually excluded at more than 2~sigma significance, when the three
sets of contours are taken into acount simultaneously (or any
combination of two of them).

To reproduce simultaneously the lack of close pairs and the clustering
below $\sim 80\hmpc$ in RIXOS, $r_0$ has to be between the lower
dashed line and the solid line in Fig. \ref{Fig7}. This translates to
$1.5\approxlt r_0\approxlt 3.8\hmpc$ for comoving evolution, $1.9
\approxlt r_0\approxlt 4.8$ for stable clustering and $2.2 \approxlt
r_0\approxlt 5.5$ for linear evolution of clustering ($>$2~sigma
significance).

The addition of the DRS limits reduces slightly the upper limit on
$r_0$ for the comoving evolution model: $1.5 \approxlt r_0\approxlt
3.3\hmpc$, leaving the others unchanged.

Using $q_0=0$ increases the limits: comoving evolution is
allowed for $2\approxlt r_0\approxlt 4.2\hmpc$, $2.5 \approxlt r_0\approxlt
5.8$ for stable clustering and $2.9 \approxlt r_0\approxlt 8.3$ for linear
evolution of clustering. These are also $>$2~sigma limits.

We have addressed the question of how much our results depend on the
exact $\rc$ value chosen to reflect the scales at which there is some
signal in our RIXOS sample. We have repeated the calculations for
$\rc\leq 43.96\hmpc$, the first comoving distance at which there is any
signal above 2~sigma. They are very similar to those for $\rc\leq
83.03\hmpc$, the lower 2~sigma limits being $\sim$19 per cent
higher. We therefore conclude that $\rc\leq 83.03 \hmpc$ is
representative of the whole $\rc\approxlt 40-80\hmpc$ range.

\section{DISCUSSION}
\label{discuss}

The clustering of galaxies seems to evolve according to the stable
model or even faster with correlation lengths $r_0\sim 3-7\hmpc$.
Smaller correlation lengths ($r_0\approxlt 3\hmpc$) would be required
if the evolution is comoving (see e.g. Infante, de Mello \& Menanteau
1996, Le F\`evre et al. 1996, Hudon \& Lilly 1996, Carlberg et
al. 1997, Brainerd \& Smail, 1998). These values are of the order of
our limits for the corresponding evolution models, and therefore our
results imply similar clustering properties of galaxies and X-ray
selected AGN.

However, clustering of optically--UV--selected AGN appears to be stronger than
that implied by our results. For instance, Croom \& Shanks
(1996) find $r_0=5.4\pm 1.1\hmpc$ and comoving evolution assuming a
$\gamma=1.8$ power law. Using a biasing model they obtain $r_0\sim
7-8\hmpc$ and slow evolution (comoving or slightly faster). These
results were obtained at $\rc\leq 10\hmpc$. They are higher than our
RIXOS and DRS upper limits at comparable scales.

A number of recent works have found an increase in the clustering of
AGN for increasing redshift: Stephens et al. (1997) found $r_0=18\pm
8\hmpc$ assuming comoving clustering evolution with a sample of 56
$z>2.7$ AGN over $\sim$22~deg$^2$. This is much higher than our
corresponding upper limits, and assuming evolution in comoving
coordinates would lead to $p<0$, again in conflict with our results.
However, we only have 5 AGN with $z>2.7$, so we essentially don't have
any information at these redshifts.  A clustering evolution scenario
in which clustering is strong at high redshifts, then it decays, and
grows again at lower redshift would in principle be compatible both
with their and our results. This is qualitatively the behaviour of the
model suggested by Bagla (1998), in which higher (rarer) mass
overdensities collapse early and cluster very strongly. The clustering
amplitude then decreases while lower and lower mass objects
collapse. When the average mass objects have collapsed, clustering
starts growing again because of their mutual gravitational attraction.

La Franca, Andreani \& Cristiani (1998) found a 2~sigma significant
increase in the quasar (AGN with $M_B\leq 23$) clustering amplitude
between $z=0.95$ and $z=1.8$ using a new sample of objects. However,
if other samples are also taken into account, the significance of this
decreases. In particular, if we use the four values for
$\bar\xi(\rc\leq 15 \hmpc,z)$ that are mutually independent in their work
and fit $\bar\xi(15,z)$ to those data, $p\geq 0$ cannot be excluded at
more than 75 per cent probability. The data points used in this fit
are: $\bar\xi(15,0.97)=0.5\pm0.2$ and $\bar\xi(15,1.85)=0.8\pm0.3$
from La Franca et al. (1998), $\bar\xi(15,0.05)=0.2\pm0.3$ from Boyle
\& Mo (1993) and Georgantopoulos \& Shanks (1994), and
$\bar\xi(15,3.1)=1.2\pm0.7$ from Kundi\'c (1997). Moreover, a 2~sigma
effect as the one reported by La Franca et al. (and indeed, as our
own clustering detection) is not very significant and requires further
work to confirm or reject it.

An interesting conclusion of our results applies to the estimates of
the anisotropies introduced in the X--ray background (XRB) by source
clustering.  Several studies have used the angular Auto
Correlation--Function (ACF) of the X--ray Background to constrain the
contribution of AGN to the XRB (Carrera \& Barcons 1992, Carrera et
al. 1993, Georgantopoulos et al. 1993, Danese et al. 1993, Chen et
al. 1994, So\l{}tan \& Hasinger 1994). In general, those studies
coincided in stating that sources clustered on scales of 6 to 8$\hmpc$
with a correlation fixed in comoving coordinates could not produce
more than about half of the XRB. A population of sources with a
smaller correlation length or faster evolution (stable or linear)
could make up the remaining XRB. These constraints are relaxed by our
work, since it appears that X-ray selected AGN present weaker
clustering and  a faster than
comoving evolution in their correlation function.

Recent soft X--ray surveys show that soft broad--line AGN only contribute
$\sim$50-60 per cent of the soft XRB, and that the contribution from
harder sources (NELGs or absorbed AGN)
grows at faint fluxes (Boyle et al. 1994, Page et al. 1996,
Romero--Colmenero et al. 1996, Almaini et al. 1996). Whatever the
ultimate nature of the hard sources turns out to be, if their clustering
properties are like those of galaxies or like those of the AGN studied
here, the whole soft XRB intensity could be produced by AGN and NELGs
with stable or linear clustering evolution and $r_0$ values within the
limits found here (see e.g. Fig. 5 of Carrera \& Barcons 1992).

It is also interesting to assess the impact of our studies on the
contribution of AGN to the hard XRB. If hard X--ray AGN cluster as the
soft X--ray AGN considered here, they could produce the whole hard XRB
without exceeding the upper limits from the ACF (Carrera et al. 1993,
Danese et al. 1993). At the moment, hard X--ray surveys only resolve a
small fraction of the hard XRB (Ueda et al. 1998).

Finally, we can use our upper limits on $r_0$ from the smallest separation
RIXOS pair to limit the ratio of bias parameters between  X--ray
selected AGN and {\sl IRAS} galaxies: ${b_X / b_I}\sim
 {\sigma_{8X} / \sigma_{8I}}$, where:

\begin{equation}
\sigma^2_8={ 72(r_0/8)^\gamma \over
2^\gamma(3-\gamma)(4-\gamma)(6-\gamma) }
\end{equation}

\noindent is the variance of the counts in a sphere of radius $8\hmpc$
for a power--law clustering model, and $\sigma_{8I}=0.69 \pm 0.04$
(Fisher et al. 1994).  Our results for the three evolution models
considered span $b_X / b_I\approxlt 0.8-1.7$.  These upper limits are
somewhat smaller than previous estimates: $b_X\sim (6.8\pm 1.6)
\Omega^{0.6}$ from the comparison of the dipole of bright 2--10~keV
selected AGN with the motion of the Local Group (Miyaji 1994),
$b_X/b_I\sim 1.5-5$ from cross--correlation of those AGN with {\sl
IRAS} galaxies (Miyaji 1994), and $b_X\approx 5.6$ from a comparison
between the X-ray and microwave backgrounds (Boughn, Crittenden \&
Turok 1998). This difference might be due to the different scales
sampled. Our result applies to scales $\sim 10\hmpc$ which is where
fluctuations are usually normalised, whilst the other studies measure
the bias parameter on much larger scales $\sim 1000\hmpc$. Our results
are however similar to those of Treyer et al. (1998)
($b_X\sim0.9-1.8$) obtained from a study of the harmonic coefficients
of the large angular scale fluctuations of the XRB from {\sl HEA01 A2} data.

\section{CONCLUSIONS}
\label{summa}

The spatial correlation function of soft X--ray selected AGN has been
studied here using data from two different X--ray surveys: the \ROSAT{
} Deep Survey (DRS) and RIXOS.

Some indication of clustering at the 2~sigma level in RIXOS in the
$z=0.5-1$ redshift range has been found using a variant of the
counts--in--cells method.  A more powerful test has also been
performed, by comparing the number of pairs of sources in each sample
with the number of pairs expected from a uniform distribution of
sources, finding a $\sim$2~sigma significant clustering signal in the
whole RIXOS sample ($\langle z \rangle=0.53$) in comoving scales
$\rc\approxlt 40-80\hmpc$. No significant detection has been found in
the DRS sample ($\langle z\rangle=1.4$) or in the RIXOS sample at
small scales. Both tests are model--independent.

Quantitative measurements of the clustering and its evolution have
been obtained from the integrated spatial correlation function
(essentially the number of pairs with comoving separations smaller or
equal than $\rc$), assuming Poisson statistics, and a power--law shape
for the correlation function, with slope $\gamma=1.8$ and correlation
length $r_0$.

Combining the limits from RIXOS ($\rc\approxlt 6\hmpc$ and
$\rc\approxlt 80\hmpc$) with the DRS limits ($\rc\approxlt 6\hmpc$),
we obtain: $1.5 \approxlt r_0\approxlt 3.3\hmpc$ for comoving
clustering, $1.9 \approxlt r_0\approxlt 4.8$ for stable clustering and
$2.2 \approxlt r_0\approxlt 5.5$ for linear evolution of clustering.
Using $q_0=0$ we obtain $\sim$20--30 per cent higher limits.

These results are compatible with the clustering
properties of `normal' galaxies, but would imply weaker clustering in
X--ray selected AGN than in optically--UV--selected ones. Our results
do not support an increase in the clustering amplitude with
redshift. However, we do not have many sources above $z>2.5$, so we
cannot rule out neither support the strong clustering above that redshift,
found by Stephens et al. (1997).

AGN (and/or NELGs) clustered like the sources studied here could
produce most of the hard and soft X--ray background without exceeding
the observed limits on its Auto Correlation Function.

The lack of very close pairs in RIXOS implies that the ratio of bias
parameters between X--ray selected AGN and {\sl IRAS} galaxies (where
there is evidence that biasing is small, $b_I\approx 1$) is $b_X/b_I
\approxlt 0.8-1.7$, somewhat smaller than previous results.

A 3 sigma detection at comoving separations $\rc\approxlt10\hmpc$
would necessitate $\sim 100-110$ `RIXOS--like' fields for comoving
evolution and $r_0=6\hmpc$. This is especially relevant for future
X--ray survey identification programmmes (such as the XMM--XID) with a
view to detect AGN clustering. It is much more efficient to use a
compact connected area (circular or elongated) for this purpose,
rather than serendipitous pointings distributed all over the sky.  In
this sense, the \ROSAT{ } All Sky Survey identification programmes
will be very useful to provide a local `anchor' for X--ray AGN
clustering studies (e.g. Zickgraf et al., 1998).

\medskip
\noindent {\bf ACKNOWLEDGEMENTS}
\medskip

\noindent Special thanks are due to Omar Almaini and his colleagues for
providing us with the positions, fluxes and redshifts of the DRS AGN.
FJC and XB thank the DGES for financial support, under project
PB95-0122. XB acknowledges sabbatical support at Cambridge to the DGES
under grant PR95-490. ACF and RGM thank the Royal Society for support. This
research has made use of data obtained from the Leicester Database and
Archive Service at the Department of Physics and Astronomy, Leicester
University, UK. We thank the Royal Society for a grant to purchase
equipment essential to the RIXOS project.

{}
{}
{}
{}
{}
{}
{}

\bsp
\end{document}